\documentclass[a4paper,12pt]{article}

\usepackage[utf8x]{inputenc}
\usepackage{graphicx}
\usepackage{overcite}
\usepackage{amsmath}
\usepackage{amssymb}
\usepackage{float}
\usepackage{rotating}
\usepackage{longtable}
\usepackage{scrextend}
\usepackage[font=small,labelfont=bf]{caption}
\usepackage{url}

\usepackage[sort&compress,numbers,super,comma]{natbib}

\usepackage{setspace}
\newcommand{\onlinecite}[1]{\hspace{-1 ex} \nocite{#1}\citenum{#1}}

\title{\bf Exploration of Reaction Pathways and Chemical Transformation Networks}
\author{Gregor N.\ Simm, Alain C.\ Vaucher, and Markus Reiher\thanks{corresponding author: markus.reiher@phys.chem.ethz.ch}
\vspace{10 mm}\\
ETH Z\"urich, Laboratory of Physical Chemistry, \\ Vladimir-Prelog-Weg 2, 8093 Z\"urich, Switzerland
}

\begin{document}

\maketitle

\begin{center}
{\bf Abstract}
\end{center}

For the investigation of chemical reaction networks, the identification of all relevant intermediates and elementary reactions is mandatory.
Many algorithmic approaches exist that perform explorations efficiently and in an automated fashion.
These approaches differ in their application range, the level of completeness of the exploration, as well as 
the amount of heuristics and human intervention required.
Here, we describe and compare the different approaches based on these criteria.
Future directions leveraging the strengths of chemical heuristics, human interaction, and physical rigor are discussed.

\section{Introduction}

For a detailed analysis of a chemical system, all relevant intermediates and elementary reactions on the potential energy surface (PES) connecting them need to be known.
An in-depth understanding of all reaction pathways allows one to study the evolution of a system over time given a set of initial conditions
(e.g., reactants and their concentrations, temperature, and pressure)
and propose derivatives of the original reactants to avoid undesired side reactions.
Manual explorations of complex reaction mechanisms employing quantum-chemical methods are slow and error-prone.
Moreover, because of the high dimensionality of PESs, an exhaustive exploration is generally unfeasible.
However, to rationalize the formation of undesired side products or decomposition reactions,
unexpected reaction pathways need to be uncovered.

Complex reaction mechanisms are ubiquitous in chemistry.
They are, for instance, the basis of transition-metal catalysis,\cite{Masters2011} polymerizations,\cite{Vinu2012}
cell metabolism,\cite{Ross2008} enzyme catalysis (e.g., Refs.~\onlinecite{Jorgensen2004} and \onlinecite{Valdez2016}),
surface chemistry (e.g., Refs.~\onlinecite{Honkala2005,Medford2014,Matera2014,Reuter2016,Baxter2017,Ha2017,Ulissi2017}),
and environmental processes\cite{Vereecken2015} and are the objective of systems chemistry.\cite{Ludlow2008}
Knowing all chemical compounds and elementary reactions of a specific chemical process is essential for its understanding in atomistic detail.
Even though many chemical reactions result in the selective generation of one main product,\cite{Clayden2001} in general,
multiple reaction pathways compete with each other. A variety of side products
may be generated by reactive species (such as radicals, valence-unsaturated species, charged reactants, strong acids and bases and so forth)
or through high-energy (electronic or vibrational) states populated by light, elevated temperature, or increased pressure.

\begin{figure}[!htb]
\begin{center}
\includegraphics[width=\textwidth]{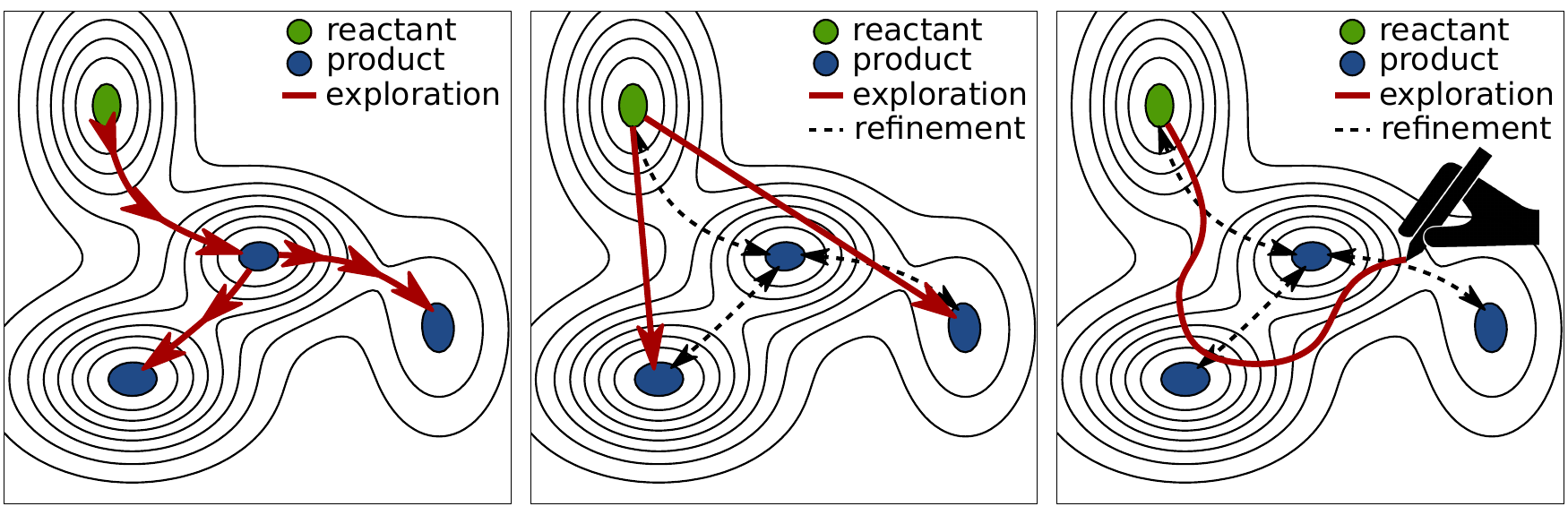}
\caption{
General strategies for the exploration of PESs.
Left: local curvature information of the PES is exploited to identify TSs and products.
Middle: through the application of heuristics new intermediates are identified.
Right: intermediates and TSs are explored interactively.
In the two latter cases of heuristics-based and interactive explorations, pathways need to be refined {\it a posteriori} to obtain MEPs.
An efficient exploration protocol may combine strategies from different classes.
}
\label{fig:approaches}
\end{center}
\end{figure}

Some excellent reviews on the exploration of reaction pathways already exist (for recent ones see Refs.~\onlinecite{Dewyer2017} and \onlinecite{Sameera2016}).
Our overview here, however, has a different focus and also extends the literature covered in previous reviews.
We organize the plethora of strategies developed for the exploration of PESs into three classes 
(some protocols combine strategies from more than one class),
which are illustrated in Fig.~\ref{fig:approaches}:

\textbf{Class 1:}
Starting from a point on the PES (indicated by the green region in Fig.~\ref{fig:approaches}, left)
new TS and intermediates are discovered by relying on local curvature information.
This process is repeated (possibly in multiple directions) until all (relevant) stationary points of the PES are explored.

\textbf{Class 2:}
Starting from a minimum energy structure, points on the PES (e.g., new intermediates or approximate TSs) are explored through the application of heuristics
(guided by chemical rules as indicated by straight lines in the Fig.~\ref{fig:approaches}, middle).
This includes, for example, the formulation of graph-based transformation rules or the application of an artificial force pushing reactive moieties upon one another.
Once a new intermediate is found, the minimum energy pathway (MEP) connecting it to the starting structure can be searched for.

\textbf{Class 3:}
The strengths of human (chemical) knowledge and intuition can be combined with ultrafast computer simulations in an interactive setting to
efficiently explore intermediates and TSs of a PES. This class has hardly been considered in previous reviews, which is the reason
why we discuss this class in some depth here.

For the accurate description of a chemical process, not only bond-breaking and bond-forming transformations are to be considered,
but also the conformational space of each intermediate must be explored.
This is particularly critical for an accurate description of thermodynamic properties (e.g., Gibbs free energies)
of compounds and for reactions in which multiple MEPs connecting the same configurational isomers can exist.
In general, all exploration strategies mentioned above can be employed to locate conformational isomers.
It should be noted, however, that in contrast to bond-breaking or bond-forming transformations, in most practical applications,
TSs between conformers are not of interest as the time scale of a reaction can be assumed to be longer
than the time required for the conformers to equilibrate.
For a recent review on conformer generation see Ref.~\onlinecite{Hawkins2017}.

In the following three sections, we review algorithms assigned to the three classes.
The approaches differ in the degree of automation possible (an aspect which is particularly critical for large reaction networks),
the amount of heuristics required, and the degree of completeness of the exploration.
Approaches involving little or no heuristics tend to explore the PES in a more systematic fashion.
However, they are often limited by computational demands.
By contrast, heuristics-driven methods can be applied to efficiently study larger chemical systems, 
although care must be taken in order not to compromise the thoroughness of the exploration by making 
strong assumptions about a system's reactivity.

\section{Class 1: Exploiting Curvature Information}
\label{subsec:curvature}

The manual construction of approximate TS candidates is slow, cumbersome, and highly unsystematic.
Recent developments therefore focused on providing more streamlined and systematic means for the location of TSs.
Ohno and Maeda exploited curvature information of the PES in a strategy called anharmonic downward distortion following (ADDF).\cite{Ohno2004,Maeda2004,Maeda2005,Ohno2006,Maeda2013}
By distorting a minimum structure in a way orthogonal to the potential energy contours, MEPs can be found.
As a result, a path explored through ADDF is very close to the one obtained from an intrinsic reaction coordinate calculation.
Repeated application of ADDF on newly explored intermediates yields all relevant MEPs.
ADDF was successfully applied to reactions of small molecules such as formaldehyde, propyne, and formic acid.
Recently, Satoh et al.\cite{Satoh2016} systematically explored conformational transitions of \textsc{D}-glucose by employing ADDF to trace only low TS barriers.

To address the limitations of ADDF, Maeda, Morokuma, and coworkers developed the artificial force-induced reaction (AFIR) method.\cite{Maeda2010,Maeda2011,Maeda2013a,Maeda2016,Sameera2016,Yoshimura2017}
AFIR overcomes intermolecular activation energies by applying an artificial force pushing the reactants together.
When optimizing under this biasing force, the maximum energy point lies close to the true TS.
AFIR has been successfully applied to the Claisen rearrangement,\cite{Maeda2013a,Maeda2016} cobalt-catalyzed hydroformylations,\cite{Maeda2013a,Maeda2016}
an aldol reaction,\cite{Maeda2016} and the Biginelli reactions.\cite{Puripat2015}
A consequence of the algorithm is that many random initial orientations of the two reactants need to be sampled before all relevant pathways are found.
In addition, human input will often be required to select pairs of reacting molecules to circumvent a combinatorial explosion
and to work with the amount of data produced.

Another systematic approach to explore potential energy surfaces starting from some configuration of a reactive system
is reactive molecular dynamics (MD) simulations, in which the Newtonian equations of motion are solved for the nuclei or atoms 
to explore and sample the part of configuration space
that is accessible under the constraints imposed by a predefined thermodynamic ensemble.
The force is either calculated as the negative gradient of a classical force-field potential energy model or of an electronic
energy (in {\it ab initio} MD) evaluated with an approximate quantum chemical method (typically from semi-empirical or density functional
theory) for the configuration for which the force is needed for trajectory propagation.
To increase the possibility of a reaction to occur, the temperature and pressure of a simulation often need to be increased
which in turn leads to the frequent occurrence of transformations that are unlikely under common laboratory conditions.

The capabilities of reactive \textit{ab initio} molecular dynamics for studying complex chemical reactions were demonstrated
with the example of the prebiotic Urey--Miller experiment.\cite{Saitta2014,Wang2014,Wang2016}
As the configuration space can become very large, comprising multiple copies of all chemical species involved in the reaction,
computational costs of carrying out first-principles calculations grow rapidly.
This issue can be overcome by the application of a reactive classical force-field (for a recent review see 
Ref.~\onlinecite{Meuwly2018}).\cite{vanDuin2001,Dontgen2015}
Unfortunately, apart from reduced accuracy, force-field parameters will, in general, not be available for any type of system which limits their applicability.
Therefore, hybrid quantum-mechanical--molecular-mechanical approaches have been frequently applied to explore
reaction pathways of complex systems with many degrees of freedom such as enzymatic reactions
(for examples see Refs.~\onlinecite{Fischer1992,Florian2003,Garcia-Viloca2004,Imhof2009,Reidelbach2016,Imhof2016} and reviews by Senn and Thiel~\cite{Senn2007a,Senn2007,Senn2009}).

Naturally, MD simulations employing classical and \textit{ab initio} force fields can be applied to sample conformational degrees of freedom.
For systems featuring many degrees of freedom (e.g., proteins) enhanced sampling techniques such as local elevation\cite{Huber1994}
or metadynamics\cite{Laio2002} are mandatory (see also Refs.~\onlinecite{Christen2008,Bernardi2015}).
Recent progress has been reviewed in Refs.~\onlinecite{Shim2011,Ballard2015,DeVivo2016}.
These approaches are among the most complex and time-consuming for conformational sampling.\cite{Tsujishita1997}
Stochastic methods based on Monte Carlo (MC) simulated annealing are often faster than MD methods.\cite{Wilson1991,Sperandio2009,Grebner2011}
By sampling low-lying eigenmodes they require less computational effort than MD simulations.
Both MD and MC approaches are computationally too expensive for a high-throughput setting.

In 2013, Liu and coworkers proposed the stochastic surface walking (SSW) method for exploring PESs.\cite{Shang2013}
Their approach is based on bias-potential-driven dynamics and Metropolis MC sampling.
Stationary points are perturbed toward a new configuration through randomly generated modes of displacement
and the subsequent construction of a biasing potential.\cite{Shang2013,Zhang2013a,Shang2014,Zhang2015}
For example, the SSW method was applied to predict the structure of complex fullerenes\cite{Zhang2013a} and phase transitions\cite{Shang2014}
and to study the hydrolysis of epoxypropane and the decomposition of $\beta$-\textsc{d}-glucopyranose.\cite{Zhang2015}

Mart\'inez-N\'u\~nez and coworkers developed an approach termed TS search using chemical dynamics simulations (TSSCDS),\cite{Vazquez2015,Martinez-Nunez2015,Martinez-Nunez2015a,Varela2017,Rodriguez2018}
in which high-energy dynamics employing semi-empirical quantum chemical methods are performed to induce reactions to occur at high rates.
Vibrational modes are populated to increase the rate at which TSs can be overcome.
For large systems, due to their large number of vibrational modes, manual intervention is required to steer the simulation into directions of interest.
The trajectory generated through the simulation is subsequently post-processed, and bond-forming and bond-breaking events are identified.
TS guesses are extracted from the trajectory and refined with semi-empirical or density-functional methods.
The TSSCDS method was successfully applied to reactions involving vinyl cyanide, formaldehyde, and formic acid and to study cobalt catalysis.

Usually, approximate TS candidates are then refined by utilizing Hessian information, i.e., second-order derivatives of the potential energy with respect to the nuclear coordinates.
The Hessian of a TS guess structure is required to obtain the normal mode representing the reaction coordinate to follow.
Eigenvector following (EVF)\cite{Cerjan1981,Simons1983,Davis1990,Wales1992,Wales1993,Jensen1995,Doye1997} is the most prominent example of such an approach,
that will be reliable if the TS guess structure is close to the true TS.
For large molecules, a full Hessian calculation will become computationally demanding or even unfeasible even if the Hessian is approximated.
Therefore, several algorithms have been developed to circumvent the calculation of the full Hessian.
A quasi-Newton--Raphson method was introduced by Broyden,\cite{Broyden1967}
in which an approximate Hessian is built from gradients only and then updated by the gradients of intermediate points obtained during the optimization.
Other approaches include schemes proposed by Munro and Wales\cite{Munro1999} that avoid the full diagonalization of the Hessian,
Lanczos subspace iteration methods,\cite{Malek2000}
and Davidson subspace iteration algorithms.\cite{Deglmann2002,Reiher2003,Sharada2012,Bergeler2015a}

Wheeler and coworkers developed a computational toolkit called AARON (An Automated Reaction Optimizer for New catalysts)\cite{Wheeler2016,Doney2016,Guan2017,Guan2018}
which constructs initial TS structures by mapping key atoms of new catalysts onto the positions of the corresponding atoms in previously computed TS structures based on a model catalyst.
AARON then executes a protocol of computations consisting of constrained and unconstrained optimizations to yield optimized TS structures.
AARON was successfully applied to bidentate Lewis-base catalyzed allylation and propargylation reactions.

\section{Class 2: Structure Hopping by Chemical Heuristics}
\label{subsec:intermediates}

Exploration strategies solely based on curvature information of one PES are often
unsuitable for large chemical systems with many degrees of freedom and for reactive systems
that require many PESs for their description owing to multiple steps with incoming reactants and outgoing side products.
Starting from a minimum structure, it can be more effective to apply heuristics to rapidly identify potential products,
and subsequently, search for a MEP connecting them.
If two endpoints of an elementary reaction are known, interpolation methods (see, e.g., Ref.~\onlinecite{Halgren1977})
and string methods\cite{Ayala1997,Henkelman1999,Henkelman2000,Henkelman2000a,Maragakis2002,E2002,E2005,Behn2011,Behn2011a,Sharada2012,Zimmerman2013,Vaucher2018} can be applied to locate the MEP connecting them.

Conceptual knowledge of chemistry can be exploited to rapidly identify potential candidates for intermediates connected to the starting structure through an elementary reaction,
in particular, if the type of reaction mechanism relevant to the reactive system under study is known.
For example, from reaction databases or chemical heuristics, transformation rules can be formulated and applied to graph representations of the reacting molecules.
These rules originate from concepts of bond order and valence and are therefore best suited for organic chemistry.
In 1994, Broadbelt and coworkers pioneered this approach by introducing a method called Netgen.\cite{Broadbelt1994}
The three-dimensional arrangement of atoms in molecules is transformed into a graph structure in which atoms and bonds are represented by nodes and edges, respectively.
This gives rise to adjacency matrices which can be manipulated by matrix operations representing chemical transformations.\cite{Broadbelt1996,Broadbelt2005}
With the derived adjacency matrices, new three-dimensional arrangements of atoms are generated.
Through repeated application of these transformation rules, new molecules are added to the list of intermediates involved in the global mechanism.
In Broadbelt's original work, elementary steps were not identified, and hence, activation barriers needed to be
estimated with the Evans--Polanyi principle.\cite{Evans1938}

In a similar spirit, Green and coworkers developed a software package called Reaction Mechanism Generator (RMG).\cite{Matheu2003,Gao2016}
Many steps were taken to overcome challenges demonstrated by the approach of Broadbelt.
In particular, kinetic parameters were estimated employing quantum chemical calculations to discard products that are likely to be reachable only by overcoming TSs featuring high activation barriers.
RMG was employed to map, in an automated fashion, the mechanisms of the pyrolysis of methane\cite{Matheu2003} and \textit{n}-butanol\cite{Harper2011} and the steam cracking of \textit{n}-hexane.\cite{Geem2005}
In addition, the Green and West groups successfully applied RMG to a variety of complex systems.\cite{Geem2005,Petway2007,Hansen2013a,Slakman2016,SeyedzadehKhanshan2016,Han2017,Dana2018,Grambow2018}
Similar to Netgen, RMG is ultimately limited by the application of concepts of bond order and valence.

Green and coworkers developed an approach in which graph transformations and path refinement algorithms are combined to find reaction pathways.\cite{Suleimanov2015,Grambow2018}
Very recently, they explored the reaction network of the simplest $\gamma$-ke\-to\-hy\-dro\-per\-oxide, 3-hy\-dro\-per\-oxy\-pro\-panal,\cite{Grambow2018}
by applying the Berny algorithm\cite{Schlegel2004,Schlegel1984,Peng1998} coupled with the freezing string method,\cite{Behn2011} single- and double-ended growing string methods, and the AFIR method.

Recently, Kim and coworkers utilized chemical heuristics to rapidly search for reaction pathways.\cite{Kim2018}
Through the application of molecular graph and reaction network analyses, they explored a so-called minimal reaction network
consisting of intermediates that can be reached from the starting structures within a fixed number of bond dissociation and formation reactions.
The minimal network is subjected to quantum chemical calculations to determine the kinetically most favorable reaction pathway.
They applied this ansatz to recover the accepted mechanisms of the Claisen ester condensation and of cobalt-catalyzed hydroformylation reactions.\cite{Kim2018}

The ZStruct approach developed by Zimmerman utilizes connectivity graphs to identify potential intermediates are reachable when connections are formed or broken.\cite{Zimmerman2013a,Zimmerman2015}
Intermediates are then subjected to a double-ended MEP search with the growing string method.\cite{Zimmerman2013b,Zimmerman2013,Zimmerman2015a,Jafari2017}
A limitation of ZStruct is the requirement that two reactants need to be aligned so that the reactive complex is close to a MEP.
As a result, this approach is best suited for intramolecular reactions.\cite{Dewyer2017}
In addition, if two intermediates are connected by two elementary reactions, ZStruct will struggle to find a TS.
Despite its shortcomings, ZStruct uncovered an unexpected side-reaction that was hampering a Ni-based C--H functionalization catalyst.\cite{Nett2015}
Furthermore, several other (catalytic) reactions have been studied with ZStruct.\cite{Li2016a,Pendleton2016,Zhao2016}
Recently, Dewyer and Zimmerman addressed the limitations of ZStruct and developed ZStruct2.\cite{Dewyer2017b}
In ZStruct2, reactants are prealigned to sample so-called driving coordinates that describe the expected elementary reactions.
ZStruct2 was successfully applied to study transition metal catalysts.\cite{Ludwig2016,Smith2016,Ludwig2017,Dewyer2017a}

In the Habershon's approach,\cite{Habershon2015,Habershon2016} connectivity graphs describe intermediates.
Reaction pathways are examined by dynamics simulations over a Hamiltonian that can be updated to suit a change in the connectivity graph.
The trajectories are processed and unique pathways are refined.
Compared to the approach developed by Green and coworkers, Habershon's method explores the potential energy surface more extensively, at the cost of running dynamics simulations.

West and coworkers demonstrated a novel approach to predict TS structures with a group-additive method.\cite{Bhoorasingh2015}
Distances between reactive atoms at the TS are estimated with molecular group values so that an approximate TS can be constructed by distance geometry (DG)
-- a stochastic method that will be discussed below.
The estimated TS structures are then optimized with standard electronic structure methods.

Aspuru-Guzik and coworkers developed a methodology based on formal bond orders\cite{Rappoport2014,Zubarev2015,Rappoport2018} to model prebiotic reactions.\cite{Butlerow1861}
Instead of specifically encoding elementary reactions, transformation rules were based on a concept popular in organic chemistry, commonly denoted as 'arrow pushing',\cite{Levy2017}
and activation barriers were estimated (in Ref.~\onlinecite{Rappoport2014} by employing Hammond's postulate).

While being computationally efficient, graph-based descriptors rely on the concept of valence which may perform well for many organic molecules
but may fail for systems containing species with complex electronic structures such as transition-metal clusters.
Furthermore, to ensure an exhaustive exploration with such an approach, completeness of the set of transformation rules is required.
However, for an arbitrary, unknown chemical system this cannot be guaranteed.
One will then be restricted to known chemical transformations, which may hamper the discovery of new chemical processes.
We, therefore, pursued a more general approach based on first-principles heuristics that is applicable to any molecular system, also those containing transition metals.
The general applicability of our approach is due to the fact that all heuristic rules are derived from the electronic wave function
that is agnostic with respect to the composition of a molecule.

In order to explore a PES we start from high-energy structures related to elementary reaction steps. For this purpose, 
molecular structures of reactive complexes are generated according to rules derived from conceptual electronic-structure theory.\cite{Geerlings2003,Geerlings2008,Proft2014}
Subsequently, they are optimized by quantum-chemical methods to produce stable intermediates of an emerging reaction network.\cite{Bergeler2015}
Pairs of intermediates in this network that might be related by an elementary reaction according to some structural similarity measure are then detected in an automated fashion
and subjected to an automated search for the connecting TS.
Compared to the other approaches, ours does not rely on the concept of chemical bonds and valence.
In contrast, \textit{reactive sites} are detected by relying on reactivity measures calculated from the electronic wave function.
This protocol has been implemented in a software called \textsc{Chemoton}
(named after a theory for the functioning of living systems proposed by G\'anti\cite{Ganti1975}).
We demonstrated the capabilities of our approach 
with the example of synthetic nitrogen fixation with the Yandulov-Schrock complex\cite{Yandulov2002,Yandulov2003}
where we were able to identify thousands of protonated structures leading to side 
and decomposition reactions that explain the catalyst's low turnover number.\cite{Bergeler2015a}
We also studied a prebiotic reaction to elucidate different pathways of early sugar formation\cite{Butlerow1861,Eschenmoser1992,Delidovich2014}
through the construction of a reaction network of unprecedented size.\cite{Simm2017a}

Our approach possesses additional salient features incorporated to make quantum chemical explorations a peer to experimental kinetic studies.
To fight the combinatorial explosion of structures, we have combined the exploration with explicit kinetic modeling\cite{Proppe2018} in order to avoid wasting  
computational resources on regions of configuration space that are kinetically not accessible under reaction conditions (in our original
implementation\cite{Bergeler2015} we already applied simple energy cutoffs for this purpose). For specific exploration problems (e.g., those
that are of purely organo-chemical nature), we can exploit fast 
semi-empirical models, for which we implemented a stand-alone program that offers energies and gradients for all major models\cite{Husch2018b}.
However, we also developed a universal semi-empirical
approach that can even deal with transition-metal complexes as it exploits the computational benefits of the neglect of diatomic differential overlap 
approximation by on-the-fly parametrization to more accurate calculations reference calculations that are then reliable 
for sequences of similar structures\cite{Husch2018}. Most important in this endeavor to exploit fast (and hence, less accurate) methods 
to the largest extent is the possibility to
assign errors to approximate electronic structure and physico-chemical models\cite{Simm2017,Proppe2017,Weymuth2018} in such a way that a seamless uncertainty
quantification is possible from the first-principles level to the rate constants and concentration fluxes of kinetic modeling\cite{Proppe2017a}. We
demonstrated that fast semi-empirical and also density-functional methods can be calibrated in an exploration 
by comparison to {\it ab initio} reference data obtained for structures determined in an automated
fashion on the fly.\cite{Simm2018} For this purpose,
we exploited confidence intervals of machine learning approaches (demonstrated at the example of Gaussian process regression).\cite{Simm2018}
It is important to note that we did not parametrize a machine learning model (which is likely to require too much
data in order to be sufficiently reliable), but instead we only exploited the feature of such models that error estimates become available.

Finally, we again need to address the problem of conformational flexibility, which opens up many routes across a PES to arrive from one compound
at another one.
There exist several efficient approaches for the exploration of conformers that are related to this second class of exploration strategies.
DG methods stochastically generate sets of atomic coordinates which are refined against a set of interatomic distance constraints.
Generated conformers are usually optimized with a molecular mechanics force field or with quantum-chemical methods to afford a candidate conformer.
Implementations of DG can be found in \textsc{DG-AMMOS}\cite{Lagorce2009} and \textsc{RDKit}.\cite{Riniker2015}

To reduce the conformational space that needs to be explored, the so-called rigid-rotor approximation is often introduced\cite{Hawkins2017}
in which bond lengths and bond angles are kept fixed so that only torsional degrees of freedom are sampled.
Genetic algorithms are a prominent class of methods for stochastic sampling of vast torsion-angle spaces as implemented in 
\textsc{Balloon\_GA},\cite{Vainio2007}
but MC algorithms have also been used for stochastic sampling of torsional angles.\cite{Leite2007,Miteva2010}
A general problem with stochastic methods is the possibility of missing relevant (i.e., accessible) conformers.
As a result, the amount of sampling required is not known.
Systematic conformer generation methods, which rely explicitly on the rigid-rotor approximation, attempt to enumerate all possible
torsional degrees of freedom of a molecule.
However, systematic enumeration of all possible torsion angles based on a starting conformation in a brute-force fashion will result
in a combinatorial explosion of candidate conformers.
Rule-based conformation generators limit the conformational space they explore.
These rules are usually derived from analyses of torsional angles in solid-state structures found in databases such as the Protein Database\cite{Hawkins2010}
or the Cambridge Structural Database.\cite{Scharfer2013,Guba2016}

We note that the generation of conformers for transition state structures
appears to be a hardly tackled problem despite its importance for subsequent kinetic
modelling. Although one may generate (or sample) conformations of some transition
state structure with the methods discussed, this could lead to different transition
state structures upon optimization. It is then not obvious, how to properly incorporate
this set of structures into an absolute rate theory calculation for an elementary
step from one compound to another. This becomes particularly difficult for conformational
freedom in the solvent surrounding a reactive system at the transition state.

\section{Class 3: Interactive Steering}

Human chemical intuition and computer simulations can form a closed feedback loop to efficiently explore particular regions of interest on a PES.
This approach does not require the formulation of heuristic transformation rules but is based on human insight and local slope information.
Interactive scientific computing very early addressed different human senses.
The first steps toward combining computer simulations and the sense of touch were conceived 
in the early 1970s.\cite{Batter1971a,Noll1972a,Atkinson1977a}
These conceptual studies explored how touch communication with computers may enhance human-machine symbiosis.
Wilson and co-workers anticipated that interactivity and the possibility to \textit{feel} simulations could be beneficial in the field of chemistry to better understand molecular structure and molecular interactions.\cite{Atkinson1977a}
Since these early days, interactive immersion into a virtual molecular world has been considered as an intruiging approach for research and teaching.\cite{lanc18,aspu18}
However, a viral spreading has not yet occurred\cite{matt18} because new designs of
hardware devices for human-machine interaction in virtual and mixed realities are still under continuous 
development. It may be anticipated that the situation changes in the future when
commodity devices become easily and cheaply available as they are required for increasingly
popular computer games.

Interactivity can be achieved by run-time modification of simulation parameters or by manipulation of atoms or molecules in the simulated system.
Immersion may often facilitate and enable interaction, and accordingly, much effort has been 
devoted to designing frameworks that are as immersive as possible, often requiring specialized hardware.
Immersion implies visual feedback and may be complemented by haptic (tactile) feedback.
On the one hand, visual feedback is often realized by a three-dimensional representation of the simulated molecular system on a (computer) screen.
More advanced setups are achieved through virtual reality or augmented reality, involving theater environments such as the cave automatic virtual environment (CAVE)\cite{Cruzneira1993a} or head-mounted displays (HMDs).\cite{Ai1998a,Prins1999}
Interactive simulations are generally steered with commodity computer mouses, three-dimensional mouses, sensors that track and record movement, or hand controllers.
On the other hand, haptic feedback is achieved with devices combining three-dimensional positional input and force feedback.
In chemistry-related applications, such devices allow for selection of atoms or molecules, move them in space, and provide a force feedback resulting from the manipulation.
Some devices not only consider three dimensions for the positional degrees of freedom, but also three dimensions for the rotation of the manipulated object.
How to achieve realistic haptic feedback in a molecular context has been studied intensively, and often depends on the targeted field
of application.\cite{Krenek2003,Lee2004,Morin2007,Daunay2009a,Bolopion2009,Bolopion2010a,Bolopion2010b,Bolopion2011}
While smooth visual immersion in virtual environments usually requires an update rate of 60 Hz,\cite{Durlach1994a} the tactile sense is more sensitive 
and requires a ten- to twenty-times increased refresh rate of about 1 kHz.\cite{Mark1996a,Ruspini1997a}
In addition to visual and haptic feedback, some applications also address the auditory sense,\cite{Cruzneira1993a,Cruzneira1996a,Ferey2009,Glowacki2014a,Arbon2018} 
and clearly voice control as an input aid could also be advantageous to enhance interactivity and immersion by reducing
latency during input and structure modification.

\subsection{Interactive Approaches in Classical Simulations}

As interactivity requires fast physical property calculations (e.g., energies and forces), molecular mechanics approaches, i.e., Newtonian
mechanics with hard-wired force fields were the first target in molecular simulations because physical quantities are easy to evaluate. A prominent example is
molecular docking. The high number of possibilities for binding sites between receptors and ligands, and the challenges of implementing automated algorithms for determining them
motivated interactive approaches to take advantage of the human ability to locate reasonable candidates.
The project GROPE\cite{Ouhyoung1988a,Brooks1990} targeted the study of interacting proteins based on electrostatic and van der Waals interactions and allowed operators to manipulate molecules with a six-dimensional haptic device to explore the interaction between receptors and ligands.
In 1997, interactive molecular docking was combined with genetic algorithms in the Stalk system in a CAVE environment.\cite{Levine1997a}
It offered the possibility to not only visualize the genetic algorithm search in virtual reality but also suspend it and move the ligand to continue the search from another conformation.
While a demonstration of a clear advantage of interactivity over traditional computational protocols
requires extensive user studies\cite{Brooks1990,Brooks2014,Connor2018a},
the usefulness of haptic feedback for molecular docking was claimed in later studies along with algorithmic and technical improvements.\cite{Bayazit2001,Nagata2002a,Laiyuen2005a,Birmanns2003a,Wollacott2007,Subasi2008a,Daunay2009a,Heyd2009,Ferey2009,Anthopoulos2014a,Iakovou2014a,Iakovou2015a,Iakovou2017a}
However, we note that especially haptic pointer devices are not completely convincing in docking studies because of i) the weak
van der Waals interactions that hardly allow one to feel an attraction between host and guest until one hits the repulsive wall of the r$^{-12}$ terms and
ii) the huge number of such weak contacts which can hardly be properly addressed with a low-dimensional input/output device.

In the 1990s, steered molecular dynamics\cite{Izrailev1999} (SMD) was developed to introduce artificial forces during molecular dynamics simulations.
This allowed for computational studies of atomic force microscopy experiments and it was soon applied to the study of rare events.\cite{Grubmueller1996a,Izrailev1997a,Balsera1997a,Isralewitz1997a}
A driving force was also Jarzynski's theorem\cite{Jarzynski1997} that relates work performed on a system out of equilibrium to the free
energy difference of the start and end points in equilibrium (although this would require a slowly acting force and extensive sampling of 
the complementary degrees of freedom).
Moreover, the artificial forces in SMD are defined in advance and cannot be altered during the simulation.
The wish to do so at runtime was one motivation behind interactive molecular dynamics.\cite{Nelson1995a,Rapaport1997a,Rapaport1997b}
First approaches consisted in changing simulation parameters intermittently or in adding user-specified forces through a graphical user 
interface.\cite{Leech1996a,Prins1999,Vormoor2001a} With increasing computational power the interactive simulations have become increasingly immersive, 
and applications soon emerged in CAVE environments\cite{Cruzneira1996a} or with specialized hardware such as HMDs\cite{Ai1998a,Prins1999} or haptic devices.\cite{Stone2001,Grayson2003}
Recent years have seen continuous efforts
to design new frameworks for interactive molecular dynamics in virtual reality\cite{Ferey2008a, Dreher2013a, Dreher2014a, Connor2018a}
and to achieve a broader applicability of interactive molecular dynamics with commodity hardware.\cite{Stone2010b, Glowacki2014a}
Also, interactive molecular dynamics was recently combined with quantum chemical methods executed on graphics processing units.\cite{Luehr2015a}
Such interactive {\it ab initio} MD, however, suffered from convergence problems of iterative orbital optimization cycles that eventually
break the immersion (see below for further discussion of such issues).

Another important application of interactive approaches in computational chemistry has been the generation and manipulation of chemical structures,
which is often considered a preparatory step for a subsequent traditional calculation because of the approximate nature of the fast method
that enables interactivity.
In 1994, Brooks and co-workers introduced {Sculpt}, a modeling system for the manipulation of small proteins with continuous energy minimization.\cite{Surles1994a}
{Sculpt} constrained bond length and bond angles but continuously minimized the potential energy due to torsion angles, hydrogen bonds, van der Waals and electrostatic interactions.
{Sculpt} allowed operators to manipulate proteins with the computer mouse and could achieve an update rate of 11~Hz for a 20-residue protein.
To ease the setup of large molecular structures, molecule editors allow for fast (real-time) structure optimization with classical
force fields or very approximate noniterative semi-empirical methods as implemented in Avogadro\cite{Avogadro} and SAMSON\cite{Samson050,Bosson2012},
respectively.
Within the SAMSON environment, Redon and co-workers applied adaptive dynamics to enable interactive modeling of larger proteins\cite{Rossi2007} and hydrocarbon systems.\cite{Bosson2012b}
They also relied on a continuous energy minimization and allowed for changes to the molecular structure (atom addition and deletion) in addition to its manipulation.\cite{Bosson2012b}
They recognized the limits of non-reactive force fields and also studied the application of semiempirical quantum chemical methods\cite{Bosson2012,Bosson2013} and extended the universal force field to support changes in the molecular topology.\cite{Jaillet2017a}

Different approaches emerged to allow for visualization of molecular structures or molecular dynamics simulations in virtual environments,\cite{Disz1995a,Akkiraju1996a,Salvadori2016a,Garcia2017a} sometimes along with functionality for the edition and manipulation of molecules.\cite{Haase1996a,Sauer2004,Norrby2015a}
In 2000, Harvey and Gingold explored the electron density of atomic orbitals with haptic devices.\cite{Harvey2000}
Comai and Mazza developed a similar tool to sense the electrostatic field resulting from quantum chemical calculations.\cite{Comai2009a}
Satoh et al.\ designed a framework for studying the van der Waals interaction of noble gases described by a classical force field
with haptic feedback.\cite{Satoh2006}
Haptic devices were later used for the interactive exploration of the solvent accessible surface of proteins.\cite{Stocks2009a}
Different studies demonstrated the possible benefits of immersive and interactive approaches in education,\cite{Sankaranarayanan2003a,Persson2007a,Sourina2009a,Bivall2011a}
and multiple virtual reality frameworks were designed for simultaneous immersion of multiple persons, allowing for collaborative environments.\cite{Chastine2005a,Nadan2009a,Hou2014a,Davies2016a,Mitchell2016a,Connor2018a}

\subsection{Interactivity for Quantum Mechanical Calculations}
\label{subsec:interactivity_qc}

Traditionally, quantum chemical studies have been prohibitively expensive to allow for interactivity.
This is due to the mathematical structure of quantum mechanical equations, which involve expensive calculations of huge numbers of integrals and
iterative solution procedures (in contrast to the straightforward and fast evaluation of an analytical formula for the energy and force 
derived from a classical force field). For this reason, quantum mechanical approaches have long been considered to be 
not amenable to interactive approaches (see, however, the next section).

\begin{figure}[!htb]
\centering
\includegraphics[scale=0.75]{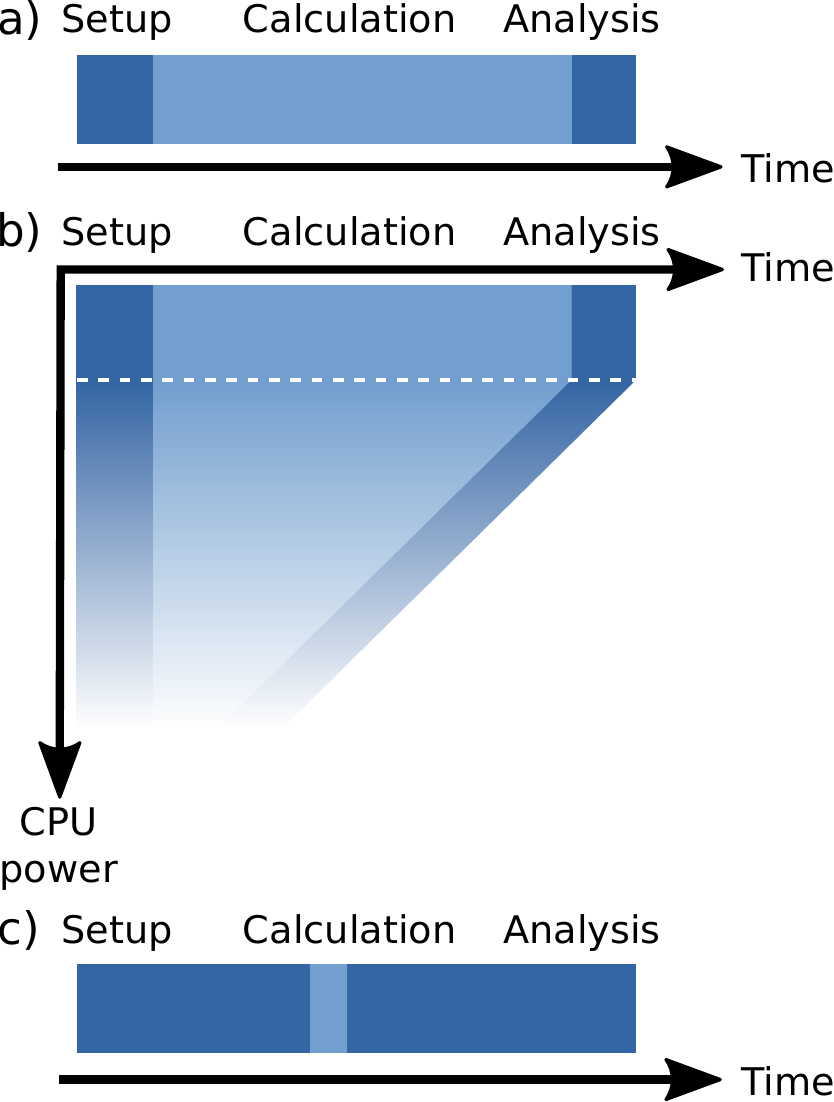}
\caption{
  Schematic course of a typical quantum chemical calculation.
  a) Traditionally, the most time-intensive step is the computation step on the CPU.
  b) With increasing CPU power, the duration of quantum chemical calculations decreases
     for a fixed system size and some given electronic structure model.
  c) For a quasi-instantaneous calculation, the most time-intensive step is now no longer the computation itself, but the
  bottleneck becomes the setup and the analysis of the results.
}
\label{fig:calculation_steps}
\end{figure}

To determine electronic energies and other properties, quantum chemical calculations require, as all computational modeling
approaches, three steps shown in Fig.~\ref{fig:calculation_steps}a:
the setup, the calculation, and the analysis of the results.
For the setup, one generally creates a text-based input file containing the calculation settings and the specification of a molecular system.
The calculation step commonly takes place on central processing units (CPUs) or cores thereof.
Quantum chemical calculations typically require computation times of a few minutes up to a few weeks.
To analyze the results, one then parses the values of interest from the generated output files.
It is known to every practitioner of computational quantum chemistry that the computation time is the most time consuming
step. In the light of the effort required for the actual calculation, setup and analysis times appear negligible.

An observation commonly known as Moore's Law states that the number of transistors on a chip roughly doubles every two years.
For quantum chemistry, this development has been highly beneficial (and has not come to end due to the growing number of cores
of a CPU and the parallelism now inherent to all quantum chemistry software).
On the one hand, it has allowed for studying larger molecular systems with increasing accuracy.
On the other hand, if the size of the studied system or the algorithm's accuracy is not increased, the calculation time will decrease over the years
benefiting from more efficient hardware that becomes available, as illustrated in Fig.~\ref{fig:calculation_steps}b.
With continuously increasing CPU power, calculation times can be greatly reduced up to the point where calculations become near-instantaneous.
In such a situation, the traditional approach (setup, calculation, analysis) is dominated by the setup and analysis, not by the calculation itself, as depicted in Fig.~\ref{fig:calculation_steps}c.
Furthermore, for near-instantaneous calculations setup and analysis are no longer separated in time.

In a traditional approach, the setup and analysis steps do not benefit from the increasing computing power.
While for time-consuming calculations the time and human effort required for setup and analysis are negligible compared to the calculation itself, this is not the case for near-instantaneous calculations.
To be able to fully benefit from near-instantaneous calculations, one must get rid of the text-based setup and analysis in the traditional 
sense altogether and consider fully interactive approaches to reactivity studies instead. Note that even most of the advanced molecule editors 
do not alleviate this problem as they help setting up inputs and inspecting results, but in general not to the degree that this 
i) is quasi-instantaneous and ii) can cope with a never-ending data influx of new results for continuously manipulated molecular
structures.
By contrast, true interactivity in quantum mechanical calculations opens new fields of research 
that combine the strengths of computers and human skills for making unprejudiced discoveries in molecular reactivity
(by contrast to hard-wired classical force fields) as we shall discuss now.

\begin{figure}[!htb]
\centering
\includegraphics[width=\textwidth]{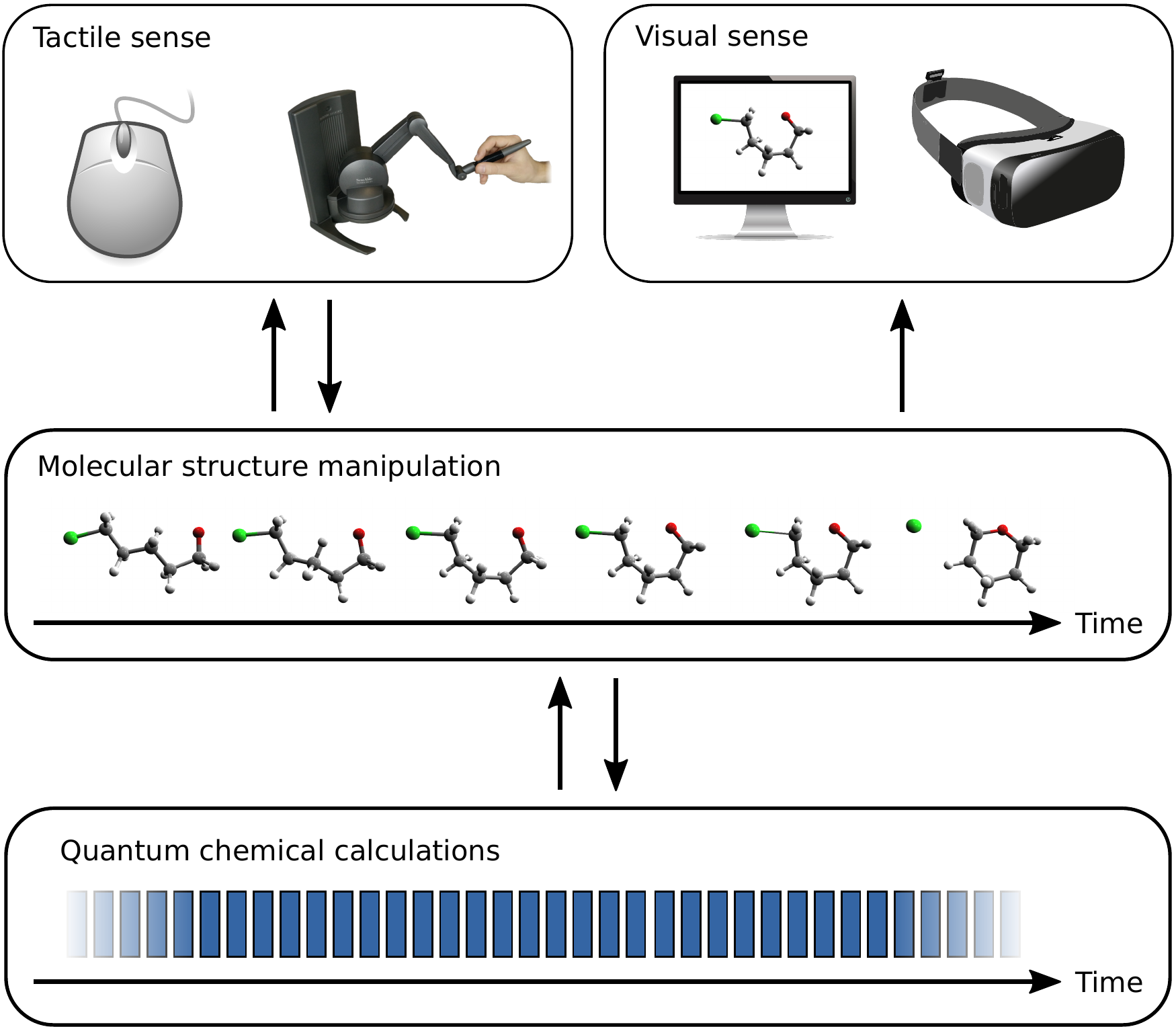}
\caption{
  Main components of a real-time quantum chemistry framework.
  The central element is the molecular structure under study (center) that is displayed (top right) to the operator.
  The operator can move atoms and experience the effects of his manipulations with a computer mouse or a haptic device (top left).
  Single-point calculations run continuously in the background (bottom) and deliver the nuclear forces underlying the reactivity exploration as well as other relevant quantum chemical properties.
}
\label{fig:rtqc_components}
\end{figure}

When considering chemical reactivity studies, we note that this naturally requires a physical modeling based on elementary particles 
and hence on quantum mechanics.
Only the well-known electromagnetic interaction operators of electrons and atomic nuclei allow us to quantify the emerging interaction energies
of atoms and fragments in arbitrary configurations of molecular systems encountered in explorations studies of chemical reaction space. 
First interactive approaches for the study of chemical reactivity appeared in the last decade.
They considered the interactive exploration of the PES based on the first principles of quantum mechanics
to determine reaction mechanisms and therefore require the application of quantum chemical methods.
In 2009, we proposed haptic quantum chemistry (HQC) as a framework to explore the potential energy surface with a haptic device
in which the force acts a descriptor for slope and, to a certain degree, also curvature.\cite{Marti2009}
The haptic force feedback first relied on an interpolation scheme that started from pre-calculated \textit{ab initio} electronic-energy data points
in configuration space.
The algorithms underlying HQC were later improved and extended for a more automated and flexible workflow.\cite{Haag2011}
Afterwards, our original HQC implementation was modified to no longer rely on interpolation, but rather on 
on-the-fly single-point calculations.\cite{Haag2013}
This new approach required a real-time quantum chemistry (RTQC) framework, which was shown to be feasible for small molecular systems within
a density functional theory model or for those consisting of about 100 atoms within semi-empirical models.\cite{Haag2014a}
A first implementation within the {SAMSON} molecule editor\cite{Samson050} with advanced state-of-the-art semiempirical methods 
demonstrated the applicability and usefulness of RTQC.\cite{Haag2014}
The main components of a RTQC framework are illustrated in Fig.~\ref{fig:rtqc_components}.

The major issue for interactivity of quantum chemical calculations is that the single-point energy
calculations require self-consistent field (SCF) iterations.
This iterative procedure creates difficulties that are already present in standard quantum chemical calculations but 
become particularly troublesome in interactive approaches.
First, one cannot predict \textit{a priori} how much time will be required for a single-point calculation,
especially for fast and drastic structure modifications introduced by the operator.
Since the execution time is linked to the number of SCF iterations, this creates a strong incentive to reduce this number,
which can be achieved by improving initial electronic densities of the SCF cycle.
Second, there will be cases in which the SCF procedure simply does not converge or requires a large number of iterations.
A nonconverging calculation results in an interruption of the flow of properties delivered to the interactive simulation, hence
destroying immersion into the virtual molecular world.
Third, there may be different solutions to the SCF equations, resulting in different electronic densities and quantum chemical properties.
In an interactive setting, convergence to an incorrect SCF solution may then lead to a biased description of the molecular system under study.

In the past years, we developed approaches that addressed all of these issues and eventually allowed us to apply SCF-based
electronic structure methods in interactive quantum mechanical settings for reaction mechanism exploration. It is important to understand that all of these 
developments, which we will discuss in the remainder of this section, address the two challenges of i) efficiency leading to faster
calculations and ii) stability ensuring reliable results. Whereas both required convincing solutions in an interactive setting where
the large frequency of results delivered (on a 100 ms to 1s rate) simply does not allow for inspecting possibly faulty results
(and therefore they must be avoided or cured at all cost), we emphasize that they also become important for automated black-box
algorithms discussed in Classes 1 and 2 above, where, due to the sheer amount of data produced, a manual check of
reliability becomes impossible (and therefore the calculations must not fail).

To address the first issue, we accelerated
self-consistent field calculations by a density propagation algorithm for sequences of (similar) structures.\cite{Muehlbach2016}
This strategy is related to other extrapolation approaches applied in \textit{ab initio} molecular dynamics.\cite{Atsumi2008, Atsumi2010} 
It delivers improved initial density matrices for single-point calculations and, hence, reduces the number of iterations
required for convergence.

Considering the second challenge, it is mandatory that visual and haptic feedback should also be available in the
case of calculations converging slowly or not at all.
In order to guarantee reliable real-time feedback, we introduced a mediator strategy.\cite{Vaucher2016}
The mediator creates surrogate potentials that approximate the potential energy surface and enable high-frequency feedback.
In addition, the mediator restricts the reactivity exploration to regions of configuration space for which the available orbitals are reliable
until new data becomes eventually available.

As for the third difficulty, incorrectly converged calculations are often ignored or go undetected even in non-interactive reactivity studies.
To ensure that interactive explorations rely on correctly converged quantum chemical calculations, we proposed a scheme which can
quickly be evaluated (as required in a RTQC framework) and which perturbs the orbitals through random mixing in such a way
that incorrect self-consistent field convergence is detected and cured.\cite{Vaucher2017}
This scheme continuously tests in the background whether lower-energy solutions can be obtained with randomly perturbed initial electronic densities.

In addition to making reactivity studies more intuitive, interactive quantum chemistry promotes novel approaches toward the way 
quantum chemical investigations are carried out.
Most importantly, the exploration need not be limited to a specific, pre-defined PES with fixed charge and spin state.
Since the quantum chemical calculations are fast, other states and properties can be calculated in the background in an automated fashion
for the structures explored to notify the operator when important facts about reactivity may otherwise go undetected.
For this scenario, we introduced the concept of molecular propensity with which multiple states of a molecular system are explored simultaneously.\cite{Vaucher2016a}
It offers chemists the possibility to discover the inclination of a reactive system to undergo (unexpected) transformations such as 
reduction, oxidation, spin crossover, photoexcitation, or protonation.

For interactive exploration of reactivity, it is imperative to keep track of the structures and reactions seen during interactive explorations.
For this, the raw exploration data must be processed to determine stable molecular structures as well as the reactions connecting them in the form of a reaction network.
This can be achieved by converting sequences of structures recorded during an exploration to B-spline curves, from which we extract candidates for stable structures and elementary reactions.\cite{Heuer2018}
The candidates are then optimized and stored in a reaction network.
Note that this network of structures can be the same data base as for the algorithms of classes 1 and 2, therefore facilitating
synergies between all classes in an integrated fashion.
To accomplish the stable optimization of reaction pathways and transition states in an automated fashion, we developed
a new double-ended method called ReaDuct that describes reaction pathways by parametrized curves.\cite{Vaucher2018}
To determine minimum energy pathways and transition states, it optimizes the parameters of the curve rather than discrete structures along the reaction path.

\section{Conclusions and Outlook}

In this work, we reviewed different approaches developed for the effective and efficient exploration of complex chemical reaction.
While approaches exploiting local curvature information of the PES are highly systematic they can be limited by computational cost,
and it is difficult to steer such an exploration to regions of a PES that are truly relevant (for a given set of reactants and
reaction conditions).
By contrast, approaches utilizing some sort of heuristics may accelerate the exploration
by employing graph-based transformation rules or concepts from quantum mechanics to efficiently discover potential intermediates.
As a result, explorations across PESs become feasible.
When heuristics are based on information directly extracted from the electronic wave function,
such heuristics can be formulated so that they are universally applicable to systems containing molecules from the whole periodic table of the elements.

Interactive approaches allow the user to guide the system of interest through the configuration space, and thus,
effectively explore particular regions that the user deems relevant to the chemical system under consideration.
However, since this type of exploration is not systematic, critical reaction pathways may remain undiscovered.
For interactivity and immersion into the molecular world, ultra-fast single-point calculations are necessary.
While such calculations are, in principle, feasible with methods of density-functional theory (DFT),\cite{Haag2013,Luehr2015a} real-time quantum chemistry frameworks have so far relied on semiempirical approaches,\cite{Haag2014a} which are also advantageous and in actual use in algorithms from classes 1 and 2 (recall,
however, our comments on reliability and error estimation in the section devoted to class 2).

\begin{figure}[!htb]
\centering
\includegraphics{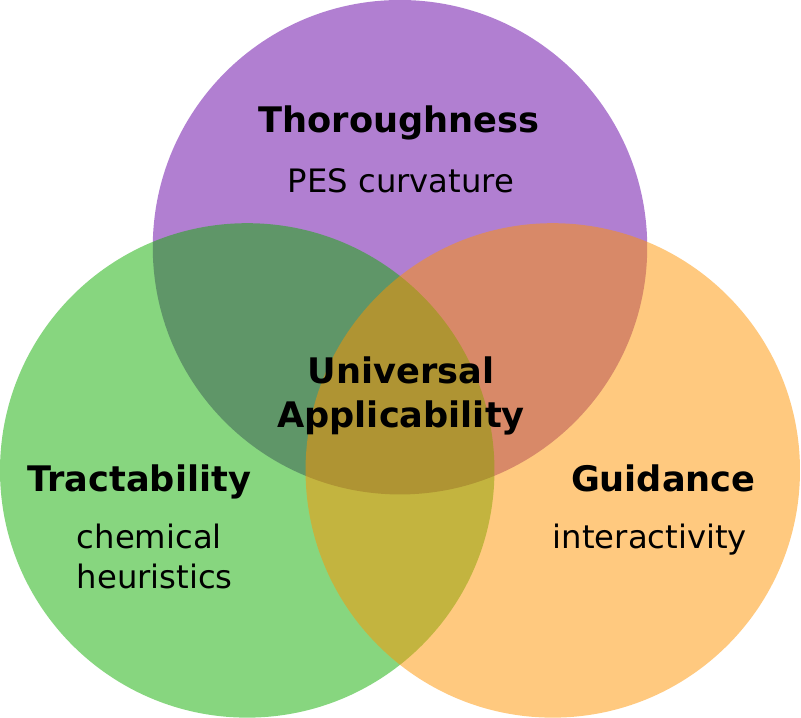}
\caption{
  The combined strengths of three classes of exploration strategies result in a universally applicable exploration protocol.
}
\label{fig:venn}
\end{figure}

We emphasize that none of the approaches in classes 1--3 can guarantee that all important elementary steps of some reaction network
are actually found during an exploration process. However, they all represent much better options than a cumbersome and limited manual
inspection providing much more detail than the latter.
It remains a challenge to strike the right balance between thoroughness and computational tractability.
In Fig.~\ref{fig:venn}, the strengths of the different classes are summarized in a Venn diagram.
Through the combination of the strengths of all three classes a truly reliable and universally applicable exploration protocol can be formulated.
In our laboratory, a software package called \texttt{SCINE}\cite{scineWeb} is currently being developed in which these three approaches are fully interoperable.

\section*{Acknowledgments}
This work was financially supported by the Schweizerischer Nationalfonds.

\providecommand{\latin}[1]{#1}
\makeatletter
\providecommand{\doi}
  {\begingroup\let\do\@makeother\dospecials
  \catcode`\{=1 \catcode`\}=2 \doi@aux}
\providecommand{\doi@aux}[1]{\endgroup\texttt{#1}}
\makeatother
\providecommand*\mcitethebibliography{\thebibliography}
\csname @ifundefined\endcsname{endmcitethebibliography}
  {\let\endmcitethebibliography\endthebibliography}{}

\end{document}